# CONSENSUS ALGORITHM FOR CALCULATION OF PROTEIN BINDING AFFINITY USING MULTIPLE MODELS

Ayşenaz Ezgi Ergin and Deniz Turgay Altılar

Department of Computer Engineering, İstanbul Technical University,
İstanbul, Turkey

## ABSTRACT

*The major histocompatibility complex (MHC) molecules, which bind peptides for presentation on the cell surface, play an important role in cell-mediated immunity. In light of developing databases and technologies over the years, significant progress has been made in research on peptide binding affinity calculation. Several in techniques have been developed to predict peptide binding to MHC class I. Most of the research on MHC Class I due to its nature brings better performance and more. Considering the use of different methods and different technologies, and the approach of similar methods on different proteins, a classification was created according to the binding affinity of protein peptides. For this classification, MHC Class I was studied using the MHCflurry, NetMHCPan, NetMHC, NetMHCCons and ssmpmbec. In these simulations conducted within the scope of this thesis, no overall superiority was observed between the models. It has been determined that they are superior to each other in various points. Getting the best results may vary depending on the multiple uses of models. The important thing is to recognize the data and act with the appropriate model. But even that doesn't make a huge difference. Since the consensus approach is directly related to the models, the better the models, the better. Xix*

## KEYWORDS

*MHC I , HLA I , binding affinity , mhcflurry , netmhcpan , netömhc , netmhccons ,ssmpmbec*

## 1. INTRODUCTION

At the population level, MHC class I genes are the locations of significant allelic diversity, with each of the hundreds of known HLA alleles linked with a stringent, possibly unique peptide-binding preference.

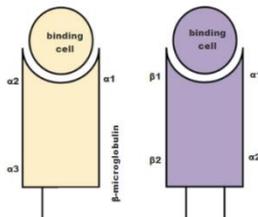

**Figure 1.1 :** Basic Structures of MHC Class I and MHC Class II

While the most selective requirement for a peptide to be delivered is a high-affinity association with MHC class I, the other steps in the antigen presentation pathway are expected to have significant secondary effects. Prediction of MHC class I-presented peptides is an important technique in vaccine development as well as research into infectious illness, autoimmunity, and cancer. [1]An immune response may be elicited and the afflicted cell may be lysed when T cells identify and bind to the peptide-MHC complex. In light of this, the binding of antigenic peptides to MHC molecules constitutes an essential stage in cellular immunity, and knowledge of the patterns underlying this event has enormous potential for use in improving human health. [2] Over the years, various methods have been used to calculate protein binding affinity, and the difficulties encountered have led to the derivation of different methods. While experimental methods were preferred at first, the expensive experimental equipment, a lot of manpower and time required by these methods, which were far from ideal, encouraged people to use computing power and technology. At first, while methods were being developed on this subject, the biggest obstacle in bioinformatics was that small data sets had difficulty in making a prediction or not being able to trust that the predictions were correct. As it is known, the data set has a very important place in calculating protein binding affinity and is directly proportional to its reliability. Although we now classify the applications that followed as score-based, it was possible to divide them into three according to the methods used in it when it was first put forward; these are imperial scoring functions, knowledge based methods and quantitative structure- activity relationships. [3] In the methods used later machine-learning methods were tried to be implemented. Overfitting was the biggest problem faced by researchers who initially saw the need to use more features due to the lack of datasets. The precision of these techniques is determined by the amount of data available defining the binding specificity of MHC molecules.

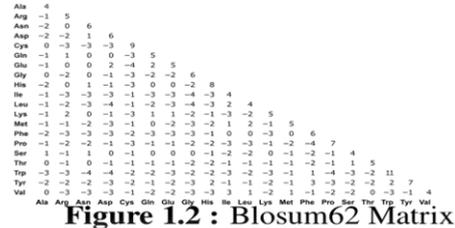

Figure 1.2 : Blosum62 Matrix

## 2. BACKGROUND

### 2.1 MHC Molecules and their General Properties

HLA is the form of the MHC gene complex that occurs in humans. However, MHC is a gene complex that expresses surface antigens in all nucleated cells in vertebrates. The main function of MHC molecules is to present antigens to T cells. It allows foreign antigens to recognize T cell against self antigens.MHC molecules convey information about the present stock of proteins within a cell to the cell surface, allowing the immune system to respond if necessary, for example, by stimulating cytotoxic T lymphocytes to destroy virus-infected cells or by activating B cells by a helper T lymphocyte. Also, MHC's three major classes are Class I, II and III. However, the main difference between HLA and MHC is their formation in vertebrates. Peptide antigen is formed as a result of peptides binding to HLA molecules and then accepted and recognized by T cells, and this is the basis of the human immune system. The amino acid sequence of the peptide binding core determines the peptide MHC binding affinity. [4]
MHC class I restricted T cell epitopes of proteins specific or over expressed in leukemia cells are potential vaccine candidates for use in immunization strategies. Class I MHC molecules feature a closed peptide-binding site on both ends and preferentially bind peptides of eight to nine amino acids that can fit within this closed binding region. [6]

MHC II molecules tend to bind to peptides of 12-20 amino acids in length, which can be considered long. The aim is to accurately predict peptide-HLA binding so that the desired response can be obtained from the immune system. Major histocompatibility complex class II (MHC-II) molecules are found on the surface of antigen-presenting cells where they present peptides derived from extracellular proteins to T helper cells. The complex formed by peptide-MHC binding is usually found on the surface of antigenic cells. Additionally, only peptides recognized by T cell receptors can trigger the immune system. Based on this, it can be said that identifying T cell epitopes may be one of the priorities for this purpose. MHC-II molecule is a heterodimeric glycoprotein consisting of an a-chain and a b-chain. Peptides presented by the MHC-II molecule are attached to a binding groove formed by residues of MHC a- and b-chains.

The peptide binding groove is open at both ends, thus allowing peptides of different lengths to be linked. Although the MHC II molecule can accommodate peptides of varying lengths, the most abundant peptides found in nature are 13 and 25 residues long. The part of the peptide ligand that interacts primarily with the MHC binding groove is called the peptide binding core and is usually nine amino acids long.

MHC Class I is easier to predict binding affinity due to its nature. Their performance is quite high in researched studies and it is generally the epitope that is studied Figure 1.1: Basic Structures of MHC more. Peptides presented by MHC class I Class I and MHC Class II molecules are mainly derived from the cytoplasm while peptides presented by MHC class II molecules are mainly derived from exogenous antigens. Peptide binding to HLA molecules is a prerequisite for peptide presentation and T cell recognition, and useful T cell epitopes contain only a subset of all HLA binding peptides. There are many factors that make it difficult to predict MHC Class II binding affinity. Polymorphic structure of the molecule, variable peptide length, PFR effect, correct peptide definition can be listed at the beginning of these. Considering these reasons and therefore all the minor factors it causes, the predictive MHC Class I binding affinity becomes much more difficult to predict. In parallel, much lower performance is observed when compared to MHC Class I binding affinity predictions.

## 2.2 Protein Binding Affinity Calculation Methods

Various models have been used that highlight the progress of computing and different features. These are examined in three different categories based on the working principle.

### 2.2.1 Score-Based Methods

The biggest difference between them is the methods used to calculate binding scores. These methods, which use more mathematics, are not suitable for all HLA Allotypes Figure 1.2: Blosum62 Matrix and assume a homogenous structure. It takes a score matrix algorithm into the center and obtains a binding score with various calculations over it, and this score outputs whether it has sufficient binding affinity or not with the threshold methods / values it has created. The most trending ones can be said to be SYFPEITHI, RankPep, PickPocket, MixMHCpred. In addition, SYFPEITHI was introduced in 1999 and we can call it the first accepted tool to calculate protein binding affinity. All of them use the PSSM matrix, as well as some of them we can come across the BLOSUM matrix as its stages. [4] It should also be added that the limitation of the scoring-based approaches is that their method of calculating prediction scores is relatively simple because they only deal with linear properties such as sequence similarity and pattern.

RankPep Peptides that bind to a given MHC molecule share sequence homogeneous attribute. Not surprisingly, sequence patterns have been traditionally utilized for the prediction of peptides binding to MHC molecules. Such sequence patterns, however, have proven to be too simple, as the intricacy of the binding motif cannot be precisely represented by the few residues present in the pattern . [7] To surmount this circumscription, RANKPEP uses Position Specific Scoring Matrices (PSSMs) or profiles from set of aligned peptides kenned to bind to a given MHC molecule as the presager of MHC-peptide binding. PSSMs for the prediction of MHC-peptide binding for a profile to be a good descriptor of the binding motif, peptides must be aligned by structural and/or sequence homogeneous attribute [8].To be more clear RankPep rates all potential fragments the width of the PSSM using the profile coefficients that score them. However, is insufficient to determine whether a peptide is a possible binder. To more precisely identify possible binders, score all of the peptide sequences contained in the alignment from which a profile is created and establish a binding threshold as the score value. [9]MHCI and MHCII molecules bind peptides in kindred yet different modes , and alignments of MHCI- and MHCII-ligands were obtained to be consistent with the binding mode of the peptides to their MHC class. RankPep can be used online via "http://imed.med.ucm.es/Tools/rankpep.html" and public avaible to everyone.

SYFPEITHI contains both MHC Class I and MHC Class II and avaible to public. Everyone can use this tool via online : http://www.syfpeithi.de/ and database is limited with published data. SYFPEITHI needed to be designed with increasing data on MHC peptide motifs, MHC ligands, T cell epitopes , amino acid sequences of MHC molecules. It can be said that it is one of the first steps taken to calculate protein binding affinity in bioinformatics after BIMAS. A two-dimensional data array is constructed, with the letters representing the row index and the pocket numbers representing the column index. Starting with the first row index, the sequence is split, and the total of the row index scores included in each molecule is computed. The method is then repeated until the conclusion of the sequence is reached. It is concluded that the integer value obtained as a result of the algorithm is suitable for binding, not suitable for binding, or has medium binding affinity. [10] Binding affinity prediction by SYFPEITHI yields a list of peptides that are presented with high probability by MHC molecules, as stated by the trustworthiness of at least 80% in obtaining the most qualified epitope, and it can be stated that the naturally presented epitope would be one of the top scoring 2% of all peptides predicted.

### 2.2.2. Machine Learning Methods

With the general development of machine-learning methods and the increase in the number of data, it has been used on protein-peptide binding affinity so that non-linear patterns can also function in calculations. Also latest developments in the application of scientific libraries utilizing machine learning techniques cleared the path for the deployment of integrated computational tools for predicting protein binding affinity. The atomic coordinates of protein-ligand complexes are used to predict binding affinity. Because of these new computational tools, researchers were able to apply a wide range of machine learning approaches to the study of protein-ligand interactions. The most important component of these machine learning methodologies is the training of a new computational model employing technologies such as supervised machine learning techniques, Convolutional Neural Networks, and Random Forest, to mention a few. [12] Machine learning-based approaches identify a peptide as a binder or non-binder by producing a score based on the retrieved representative characteristics using the training model. Although different methods and algorithms are followed, machine learning-based protein binding affinity calculation tools take place in four basic steps. [13]

- First step to obtain a train dataset composed of alleles and peptides that have previously been experimentally validated (considered a fact to bind). It is important to find verified allotypes before making a prediction at this step.

- Establishment of feures suitable for the behavior of the peptide and HLA allotypes used. It should be noted that peptides and HLAs can behave very differently. It should be noted here that while peptides have a separate nature, HLA molecules also have a separate nature. In addition, dual behaviours can also differ in this way. Feature selection should be done wisely and accurately when estimating binding affinity. In general, this step requires some biological knowledge, and over the years, many researchers have observed the biggest challenge of doing multidisciplinary research on this subject. For example, the MHC Class I molecule shows a strong bond for the same peptide, while the MHC Class II molecule can form a very weak bond.

- After the train dataset and the necessary feature coding are done, the best fit machine learning algorithm should be selected and the model created with this algorithm should be trained. These can be traditional machine learning algorithms such as RELIEF algorithm, support vector machines, linear regression or More advanced machine learning approaches, such as convolutional neural networks (CNNs), deep belief networks (DBNs), and self-encoding neural networks, may also exist. CNN is one of the easiest deep learning algorithms to apply and generalize in disciplines other than computer science. It is the most often utilized method. CNN is a type of neural network that uses feed forward. It is comparable to traditional ANN. [14]

- Optimization of the model and performance evaluation should be made. One of the biggest problems here is over fitting. If our model starts to memorize the data set we use for training more than necessary, or if our training set is monotonous, the risk of over fitting is high. If we show our test data to this model, where we get a high score on the training set, we will probably get a very low score. Because the model memorized the situations in the training set and searches for these situations in the test data set. Very bad prediction scores can be obtained in the test data set, since no memorized cases can be found in the slightest change. We can use various algorithms to avoid over fitting. The most used algorithm when calculating protein binding affinity estimation is back propagation. [15]

NN-based tools are studied in this research. In general, the NNs in the HLA-peptide-binding prediction model feature a layered feed-forward design. In a typical multi-layer feed-forward NN, the layers are composed of the input, hidden, and output layers. Each layer can comprise neurons or units that represent the signal.

NETMHCII's first characteristic is being an allele-specific method. It can only predict binding affinities in train data set and trained one by one for each MHC molecule. This methos , for being alle-specific, must be preffered when avainle dataset size is big. [16] The peptide and MHC pseudo-sequence were encoded using the BLOSUM50 matrix, and the PFR was encoded using the average BLOSUM scores at a maximum window of three amino acids at both ends of the binding nucleus. NetMHCII uses 50 networks, picking the top 10 and then applying a consensus algorithm between them . [17]

NetMHCIIPan NetMHCIIPan is a pan-specific technique for predicting any MHC molecule given a known protein sequence. NetMHCI- IPan achieves pan-specificity by combining information about the MHC-II molecule using a pseudo-sequence of residues considered to be crucial for peptide binding. NetMHCIIPan results in a universal network. Performance of pan-specific regards on nearest neighbour. [16] [18]

## 2.3. Consensus Methods

The consensus techniques' simplicity and robustness are critical. When several techniques are combined, it implies that if they perform considerably better in the evaluated conditions alone than the individual methods, they are included in the final method's definition. [19]Furthermore, it has been proven that consensus techniques, which are defined as combinations of two or more separate methods, resulted in increased prediction accuracy. Because there are so many ways, it might be difficult for a non-expert user to select the best method for predicting binding to a certain MHC molecule. As the name suggests, these methods are about combining several methods and making the decision with a consensus algorithm. The biggest reason why there are so many tools and all of them are used is that they stand out in terms of performance according to the diversity and distribution of the data. The prominent consensus models here aim to use many tools and obtain more reliable results. This is actually the general principle of consensus algorithms, which aims to conform to the majority decision. Consensus algorithms not only agree with a majority of the votes, but also accept someone who benefits them all. So, it is always a win for the network. The fact that this principle of thought has gained a lot of ground in parallel computations also suggests the hypothesis that consensus methods can increase their performance by parallelizing them.Although it is very close to parallelization, attention should be paid to feed-forward architectures. I anticipate that a distributed architecture can be used instead of a layered architecture and thus an acceleration in performance can be achieved. For example, NetMHCCons uses 3 different (2 nn-based, 1 PSSM based) methods to use the appropriate HLA allotype. This application also benefits from the nearest neighbor detection. [13] In another study, it was observed that when NetMHCII and NetMHCpan ,used as two parallel methods, which differ in their internal properties, were combined, they achieved a higher performance than both of them. [16]

NetMHC is an artificial neural network-predicated (ANN) allele-concrete method which has been trained utilizing 94 MHC class I alleles. Version 3.4 is utilized as a component of NetMHCcons1.1. NetMHCpan is a pan-concrete ANN method trained on more than 115,000 quantitative binding data covering more than 120 different MHC molecules. Version 2.8 is utilized as a component of NetMHCcons-1.1. PickPocket method is matrix-predicated and relies on receptor-pocket kindred attributes between MHC molecules. It has been trained on 94 different MHC alleles. In the PickPocket version 1.1, the matrices of pocket-library are engendered utilizing the SMMPMBEC method. NetMHCcons 1.1 server can engender presages for peptides of 8-15 amino acids in length. It withal gives a possibility to cull several lengths. Two submission types are handled - the list of peptides or a protein sequence in FASTA format. The server provides a possibility for the utilizer to cull MHC molecule in question from a long list of alleles or alternatively upload the MHC protein sequence of interest. The utilizer has a cull of setting the threshold for defining vigorous and impuissant binders predicated on prognosticated affinity (IC50) or percentage Rank. Vigorous and impuissant binding peptides will be betokened in the output. The output can withal be sorted predicated on predicting binding affinity as well as filtered on the utilizer-designated thresholds. [19]

## 3. WORKFLOW

Although models calculating protein binding affinity adopt different principles, it is impossible for them to produce a value that differs from each other because they generally have to conform to a context. Also, when labeling binding strengths (strong, ntermediate, weak) it is not expected that very different decisions will be common in heir joint decisions. Both numerical estimations and labeling studies were performed with four different models.

Model 01-09; in this study, the interquartile range is between 0.1 and 0.9 quantiles of the data. The remaining estimates were ignored and the mean value of this cluster was calculated.
Model 02-08; in this study, the interquartile range is between 0.2 and 0.8 quantiles of the data. The remaining estimates were ignored and the mean value of this cluster was calculated.

Model 035-065 ;in this study, the interquartile range is between 0.35 and 0.65 quantiles of the data. The remaining estimates were ignored and the mean value of this cluster was calculated. The biggest problem in this model is that no calculations could be made when the data did not show a normal distribution.

Model Median ;in this model, after all calculations were made, the median of the set of predicted values was created.

## 3.1 Models Used

All the knowledge gained in theory and practice shows that there is no single good model. The pros and cons of the models relative to each other directly affect the results. As a result of the researches, it is seen that consensus models perform better than other models. The biggest reason for this is that the models that seem inadequate alone have started to give better results when combined with each other. Our goal is to get the most accurate result by working together with the models. When working together, models are used with a coefficient according to the incoming sequence or not used at all. For example, the relationship between the NetMHC train dataset and the sequence directly affects the result. By looking at the Train dataset, it is calculated whether the model can be used or how much effect it should have on the result if it will be used.

### 3.1.1 Mhcflurry

The program uses a new architecture and peptide encoding technique to create allele-specific neural networks. MHCflurry beat the conventional predictors NetMHC 4.0 and NetMHCpan 3.0 in a benchmark of ligands discovered by mass spectrometry when trained on affinity measurements, notably on non-9-mer peptides. On a limited benchmark of affinity measurements, the published predictor exhibited competitive accuracy with established tools, including the recently released NetMHCpan 4.0, using mass spectrometry datasets for model selection. The prediction speed of MHCflurry topped 7,000 predictions per second, which is 396 times quicker than NetMHCpan 4.0. [21] MHCflurry creates distinct predictions for MHC allele-dependent and allele-independent effects. On existing MHC class I ligand data, including affinity measurements and MS datasets, the algorithm first trained a new pan-allele MHC class I as distinct predictors for MHC alleledependent effects. One of several design choices intended to limit the for MHC alleledependent effects predictor's tendency to learn individual antigen processing predictions signals is the use of in vitro affinity measurements in the training data, which are largely independent of individual antigen processing predictions. [22]

MHCFlurry is an open-source tool and is very simple to use. Everyone can access it at https://github.com/openvax/mhcflurry, and all of them are coded in Python, the source code is made available to the public. Class I peptide/MHC binding affinity prediction is implemented by MHCflurry. The latest version includes pan-MHC I predictors that work with any known MHC allele. MHCflurry uses the tensorflow neural network library and runs on Python. [22] The intensity of the binding relationship between a single biomolecule and its ligand/binding partner is referred to as binding affinity. The equilibrium dissociation constant (KD), which is used to quantify and rank order the strengths of biomolecular interactions, is commonly used to measure and report binding affinity. The lower the KD value, the higher the ligand's binding affinity for its target, and the higher the KD value, the weaker the attraction and binding between the target molecule and the ligand. [23]MHCFlurry provides an estimated affinity value as output. The affinities (KD) in nM are returned as predictions. The 5-95 percentile forecasts across the models in the ensemble are given by the anticipated low and prediction high fields. [21] As it can be understood from this definition, there is an inverse proportion between the affinity value and the bond.At this stage, mhcflurry-binding-affinity and mhc-binding-affinity-percentage of 3,465 8-long peptides with HLA-A0201 and HLA-A0301 alleles were calculated with this tool, which is very simple to use from the command line. When clustering the results obtained, mhcflurry-affinity-percentile was evaluated as weak between 70-100, medium between 45-70, strong bond between 0-45. Some of these data are presented in the table below. A huge number of randomly selected peptides are tested for that allele, and the mhcflurry affinity percentile provides the percentile of the affinity prediction among them (range 0 - 100). Stronger is lower. It is typical to employ a threshold of two percent. [21]

MHCFlurry is an open-source resource. The source code and implementation can be accessed at https://github.com/openvax/mhcflurry. Default features are used in this study. It was examined on various peptides and alleles as inputs without any parameter changes.

### 3.1.2 NetMHC

The first characteristic that stands out is that this model consists of 50 neural networks. The Model uses the top 10 models with the best results among 50 NNs and selects these 10 models with Pearson Correlation. The similarity metric used in protein binding affinity directly affects the result. It uses the BLOSUM62 similarity matrix, which is also used in other models, and achieves the correct result with sparse encoding. [17]The length preferences learned by NetMHC at the individual allele level approximately reflect the distribution of the data used to train the neural networks. According to one prior study, when both HLA-A*02:01 and HLAB*07:02 are measured binders, around 30% of the 9mer and 10mer peptides in the data set are measured binders, however relatively few 8mers and 11mers have a significant measured affinity for these molecules. [24] If we briefly summarize the features of the model, it makes predictions using an artificial neural network. There are 81 different human and 41 different animal alleles in the training dataset. The model can predict any length, but recommends 9 and no more than 11 amino acids. Because it is alle specific, heeding this warning is crucial for accuracy. [25]

NetMHC allows academic downloads from https://services.healthtech.dtu.dk/service.php?NetMHC-4.0. .2.2.1 Implementation It provides an e-mail address for non-academic users to communicate with. It also provides direct output with an online server for users who want to use it. In addition, the data-set can be downloaded in the same way.NetMHC Linux version 4.0 is used. In this study, progress was made with the IEDB dataset. Dataset is available at http://tools.iedb.org/mhci/download/.

### 3.1.3 NetMHCpan

Using artificial neural networks, the NetMHCpan-4.1 service predicts peptide binding to any MHC molecule of known sequence. More than 850,000 quantifiable Binding Affinity and Mass-Spectrometry Eluted Ligands peptides were used to train the approach. BA data includes 170 MHC molecules from humans and different animals. The EL data set includes 177 MHC molecules from humans. Furthermore, by uploading a full length MHC protein sequence, the user can obtain predictions for any custom MHC class I molecule. For peptides of any length, predictions can be made. [2] Since the Pan-specific approach depends on the algorithm's capacity to recognize similarities in the relationship between peptides and MHC sequences, it is obvious that the method should work better when the MHC molecule query is represented by MHC molecules closely associated with features binding specificity. [26]The panspecific method depends on neural networks' capacity to comprehend broad characteristics of the interaction between peptides and HLA pseudo-sequences in terms of binding affinity. The neural network learning process can only be aided by interactions that are polymorphic in the training set of data. Due to the lack of polymorphism at the extended MHC positions in the training data, the NetMHCpan approach would not be able to extract information from such extended pseudo-sequence mappings.

NetMHCPan allows academic downloads https://services.healthtech.dtu.dk/service.php?NetMHCpan-4.1. It provides an e-mail address for non-academic users to communicate with. It also provides direct output with an online server for users who want to use it. In addition, the data-set can be downloaded in the same way. In this study, progress was made with the IEDB dataset. Dataset is available at http://tools.iedb.org/mhci/download/.

### 3.1.4 NetMHCCons

When we look at the models in general, it is seen that all of them have advantages and disadvantages compared to each other. Therefore, it is aimed to highlight the advantages and minimize the disadvantages with the use of consensus models. Consensus models achieve results using several models. In this model, 3 different approaches are used. These are netmhcpan, netmhc and pick pocket. PickPocket is a sequence-based model, that is, it proceeds through linear features.

All three approaches and their combinations can be used to specify the consensus method when an allele is present in the training set of data. For the purpose of determining the best consensus technique, certain approaches are more crucial than others since they frequently provide the maximum performance.

NetMHCCons allows academic downloads https://services.healthtech.dtu.dk/service.php?NetMHCcons-1.1. It provides an e-mail address for non-academic users to communicate with. It also provides direct output with an online server for users who want to use it. In addition, the data-set can be downloaded in the same way. In this study, progress was made with the IEDB dataset. Dataset is available at http://tools.iedb.org/mhci/download/. Since this model is a consensus model, it is also necessary to load the PickPocket, NetMHC and NetMHCCons models it uses.

### 3.1.5 SSMPMBEC

An amino acid similarity matrix called PMBEC was created for the peptide. The matrix's inability to enable the replacement of amino acids with opposing charges is one of its distinguishing characteristics. [28] This is most likely a characteristic of the peptide in general. The PMBEC matrix was used to approach the SSM model, which originated with this notion. There are no appreciable differences in the ranges between studies using NetMHC that have been conducted for a while.

Implementation SSMPMBEC is also avaible on https://github.com/ykimbiology/smmpmbec.

## 4. RESULTS

The results clearly reveal that no general model can be chosen. The data set used in selecting the model is one of the most important considerations. When used together, the affinity calculation result is more satisfactory. The dataset used here is the IEDB training dataset. The fit of the predictions to the model is shown with various figures. As expected, the results with the most divergence occurred with 11-mer peptides. 9-mer and 10-mer peptides are more compatible than others. It has also been noticed that every model has a margin of deviation. There is a growth around the real model in the models used by removing the created outliers. This shows us that these methods are shaped around real results. The data in the graphs are shown by normalizing.Since the models created come to a conclusion in line with the models used, the accuracy of the modules used directly affects the models created. In general, the problem of incorrect labeling has occurred in the limit values.

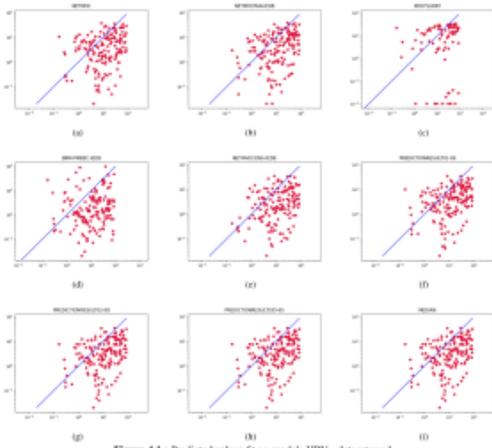
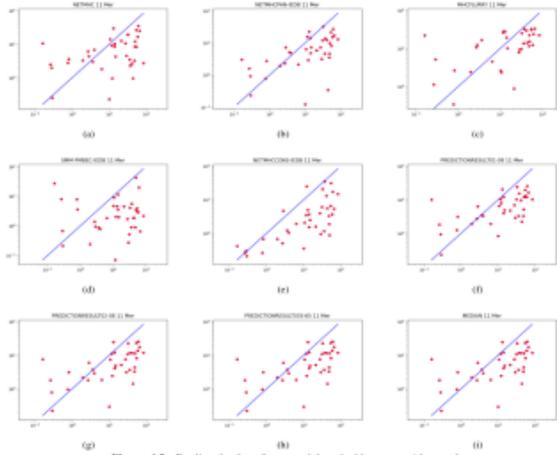

Figure 4.1 : Predicted values fit on models HPV - dataset used

Figure 4.2 : Predicted values fit on models only 11-mer peptides used

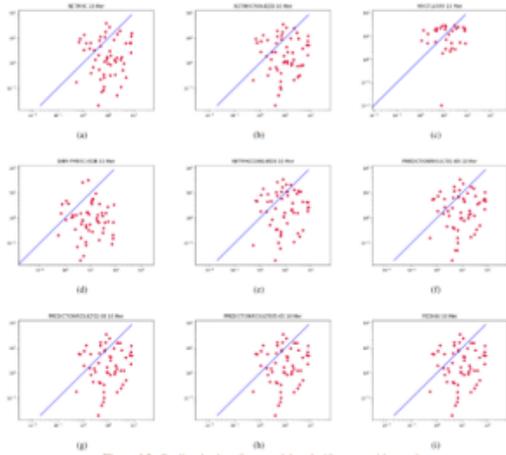
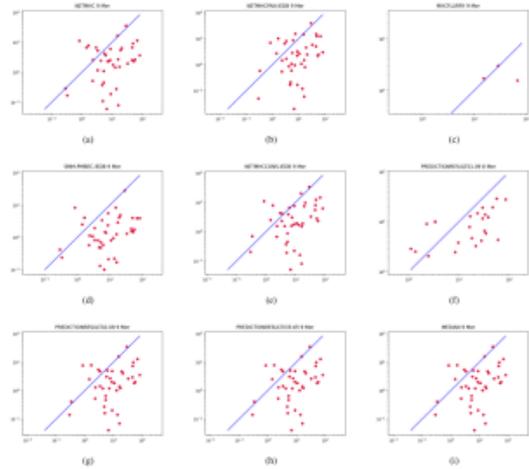

Figure 4.3 : Predicted values fit on models only 10-mer peptides used

Figure 4.4 : Predicted values fit on models only 9-mer peptides used

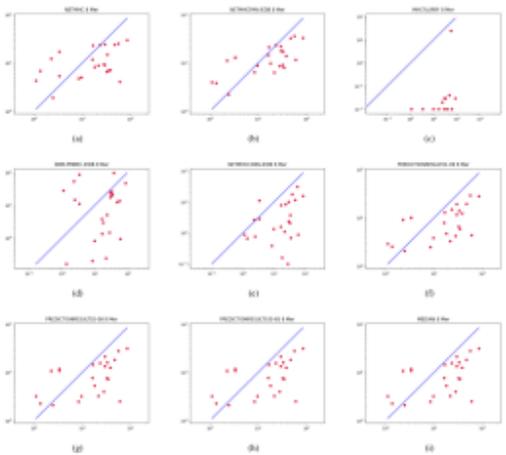
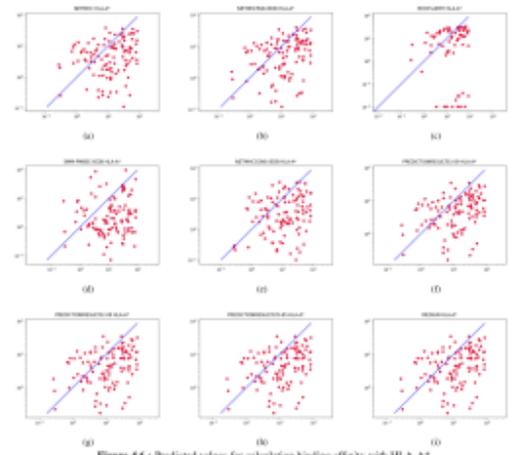

Figure 4.5 : Predicted values fit on models only 8-mer peptides used

Figure 4.6 : Predicted values for calculation binding affinity with HLA-A*

## 5. CONCLUSIONS

We cannot classify the models used as efficient or inefficient. Although consensus models give better results in general, the given data and training set directly affect the results. The biggest obstacle in this study is the limited data and limited current resources. There is definitely a data entry where each model approaches the best, and even when we compare it, there are not big differences in all of them. Since the deviation in the results of MHCFlurry and SMMPMBEC is higher, when the data set of these two models is known, using other models when the data set is more uncertain can yield healthier results.

The fact that the data is directly related to improving the models actually gave rise to the idea that the data arrangement used while improving is more important. For these inferences from scoring matrices. It is aimed to improve the model by creating scoring matrices with deep learning.

The results clearly reveal that no general model can be chosen. The data set used in selecting the model is one of the most important considerations. When used together, the affinity calculation result is more satisfactory. The dataset used here is the IEDB training dataset. The fit of the predictions to the model is shown with various figures. As expected, the results with the most divergence occurred with 11-mer peptides. 9-mer and 10-mer peptides are more compatible than others. It has also been noticed that every model has a margin of deviation. There is growing around the real model in the models used by removing the created outliers. This shows us that these methods are shaped around real results. The data in the graphs are shown by normalizing. It was tried to make sense of the data and to see the relevance of the model to the data by examining the data in different clusters. For the first clustering, the data set was divided into 4 parts according to the allele length. It was examined how much the 8-mer, 9-mer, 10-mer and 11-mer clusters overlapped the actual values with the model values. The situation of the models used and the approach created with the whole data set was examined. Figure 4.1 was created before the entire dataset was parsed. The most striking in the models may be that the MHCFlurry either converges to the most accurate or finds a value that deviates from it quite directly. While ssm-pmbec gives the farthest results in the examined models, the model closest to the real findings is 27 seen as NetMHCCons. Consensus models created show performance in proportion to the models used. While it is expected to see improvement as the outlier ratios change in the models, the best results are seen in the 01-09 model. The reason for this is that the results seen as outliers in the models have affected the results in the actual model approximating the reality. In other words, the data in the 01-09 model, which are considered outliers in other models, played a role in improving the results, that is, it helped to reduce the margin of error. Although the results in MHCflurry are very close or very far from the model, it is observed that the result is farther from the truth as the allele length decreases in clusters. The expectation for the best results was the 9-mer or 10-mer ones. In the preliminary research, it was suggested that the models were generally more successful in the 9-mer and 10-mer ones. Since the models were more successful in the predictions made for the 10-mer when compared to the 9-mer and 10 mer, it directly affected the consensus approach proposed in this research, causing the predictions for the 10-mer to obtain more convergent results. The most conspicuous is the MHCFlurry results. The results show that the model alone can give very misleading results because it is too close or too far away. Although the models using HLA-A* as input are perceived to show better results due to the fact that HLA-A* is higher in the data used, there is no big difference between the two when viewed on the basis of deviation